\documentclass[preprint,12pt]{elsarticle}

\usepackage{amssymb}
\usepackage{amsmath}

\journal{Physica B: Condensed Matter}

\begin{document}
\biboptions{sort&compress}
\begin{frontmatter}



\title{Synthesis and anisotropic magnetism of single crystalline GdPt$_2$Si$_2$}

 \author[label1]{Gustavo Gomes Vasques}
 \affiliation[label1]{organization={Centro de Ciências Naturais e Humanas (CCNH), Universidade Federal do ABC (UFABC)},
             addressline={Avenida dos Estados, 5001},
             city={Santo Andre},
             postcode={09280-560},
             state={São Paulo},
             country={Brazil}}

\author[label1]{Mateus Dutra}

\author[label1]{Pedro Caetano Sabino}

\author[label1]{Juliana Gonçalves Dias}

\author[label1]{Julián Andrés Munévar Cagigas}

\author[label1]{Adriano Reinaldo Viçoto Benvenho}

 \author{Marcos A. Avila\corref{cor1}\fnref{label1}}
 \ead{avila@ufabc.edu.br}
 \cortext[cor1]{Corresponding author}

\begin{abstract}
Single crystals of GdPt$_2$Si$_2$ were grown using the Sn flux method, crystallizing in the CaBe$_2$Ge$_2$-type tetragonal structure with space group \textit{P}4/\textit{nmm}. 
Electrical resistivity, specific heat, and magnetization data revealed the presence of a double magnetic transition with $T_N \approx 8.4~$K and $T_0 \approx 6.8~$K.
Analysis of the specific heat data suggest amplitude-modulated and equal-moment antiferromagnetic orderings, respectively. 
Field-induced magnetization and magnetic susceptibility data show a metamagnetic transition in the $H \parallel a$ direction at 2~K, as well as the suppression of the magnetic transition located at $T_0$ with increasing external magnetic field.
Electron Spin Resonance (ESR) shows the Gd$^{3+}$ resonance followed by a small second resonance. 
Peak-to-peak linewidth ($\Delta H_{pp}$) analysis reveals slight broadening at $T \sim 120$~K, indicating an increase in magnetic fluctuations at high temperatures.
Ferromagnetic (FM) local polarization at high temperatures is also observed through the $g$-factor analysis, which shows a notable positive shift ($\Delta g$).
Our results establish the fundamental physical properties of this material to aid in further understanding of the magnetism in the RPt$_2$Si$_2$ series and related non-centrosymmetric systems.
\end{abstract}







\end{frontmatter}


\section{Introduction}
\label{sec1}

Gd-based compounds are often used as a reference to understand exchange interactions in series of materials where the magnetism arises solely from the rare-earth element. 
This is due to the half-filled $4f$ shell of Gd with $S = 7/2$ and $L = 0$, implying low magnetocrystalline and negligible crystal electrical field (CEF) effects. 
The magnetic order that arises in such compounds is commonly explained by Gd-Gd dipole interactions \cite{RNiSi3_2018,RCo2Ge2_2014,RNiGe3_2010,Gd_dipoles_2003, GIGNOUX1991_rare}.
The GdT$_2$X$_2$ compounds (T = transition metals and X = Si or Ge)
generally crystallize in the ThCr$_2$Si$_2$-type structure, with a body-centered tetragonal lattice \textit{I}4/\textit{mmm}, and many of them show a complex magnetic phase diagram with multiple orderings \cite{SZYTULA_compendio_1989,Shigeoka_GdPd2Si2_2011,DUONG_GdNi2Ge2_2003, Malik_1998}. 
Recently, the widely studied GdRu$_2$Si$_2$ \cite{Khanh_GdRu2Si2_2020}, GdRu$_2$Ge$_2$ \cite{Yoshimochi_GdRu2Ge2_2024}, GdRh$_2$Si$_2$ \cite{Güttler2016} and GdIr$_2$Si$_2$ \cite{GdIr2Si2_2021} compounds have regained attention due to the successful synthesis of high-quality single crystals, which made it possible to study phenomena such as skyrmion formation in a centrosymmetric lattice and spin Rashba splitting.

\begin{figure}[]
\centering
\includegraphics[scale=1.6]{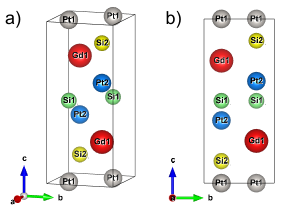}
\caption{\label{fig:structure} a) Crystal structure representation of GdPt$_2$Si$_2$ and b) its $bc$-plane projection where the absence of inversion symmetry of the Gd-ion is evidenced.}
\end{figure}

In contrast with the ThCr$_2$Si$_2$-type structure, GdPt$_2$Si$_2$ and a few others adopt the less-frequent tetragonal, non-centrosymmetric CaBe$_2$Ge$_2$-type structure (\textit{P}4/\textit{nmm} space group represented in Fig.~\ref{fig:structure}). 
All reported Gd compounds in this structure show antiferromagnetic order (AFM) followed by an additional magnetic transition below $T_N$ \cite{HIEBL1985,KACZMARSKA199581,KACZMARSKA1993151,KACZMARSKA1995208,HULLIGER1994263}. 
Moreover, only a few were reported in single crystalline form and have had their magnetic properties studied in detail \cite{GIGNOUX1991, KLIEMT2022_GdIr2Si2}.
In this work, we report the growth of single crystals of GdPt$_2$Si$_2$ by Sn flux and a comprehensive study of its magnetic, thermal and electrical transport properties through specific heat, resistivity, magnetic susceptibility and electron spin resonance (ESR) measurements.
The collective data set was used to construct the anisotropic magnetic phase diagram for this compound with $H \parallel a$ and $H \parallel c$.

\section{Experimental techniques}

Single crystals of GdPt$_2$Si$_2$ and its non-magnetic reference compound LaPt$_2$Si$_2$ were grown by the Sn-flux method. 
The starting materials were Gd (99.9\%)/La (99.995\%), Pt (99.95+\%), Si (99.999\%) and Sn (99.999\%), mixed with molar ratio of 1:3:1:95 for the Gd compound and 1:2.5:1.5:95 for La, then inserted and sealed in quartz ampoules under vacuum. 
The ampoules were heated up to 500~ºC, dwelled for 1~h, then heated up to 1180~ºC in 3~h and dwelled for 24~h to ensure a homogeneous melt of the reagents. 
Subsequently they were cooled at a constant rate of 2~ºC/h down to 680~ºC, where the ampoules were removed from the furnace and immediately centrifuged to remove the excess of Sn. 
HCl was later used to remove the remaining flux from the crystals, which have typical dimensions of $0.9\times 0.4\times 0.08$~mm$^3$.
The crystal structure was confirmed through powder x-ray diffraction (XRD) on crushed crystals using a STOE STADI-P diffractometer with Cu-K$\alpha$ radiation. 
The composition of the crystals was checked by energy dispersive x-ray spectroscopy (EDX) on a JEOL JSM-6010LA, coupled to a scanning electron microscope (SEM). 
Magnetic measurements were made on a Quantum Design MPMS3 SQUID-VSM. Four-probe resistivity and heat capacity were carried out on a Quantum Design PPMS EverCool III system.
Electron spin resonance (ESR) measurements were performed on a BRUKER - ELEXSYS 500 CW spectrometer with a TE102 cavity, using frequencies close to $\nu=9.4$~GHz (X-band) with continuous nitrogen gas flow.

\section{Results and discussions}

\subsection{Crystal structure}

\begin{figure}[ht]
\centering
\includegraphics[scale=0.3]{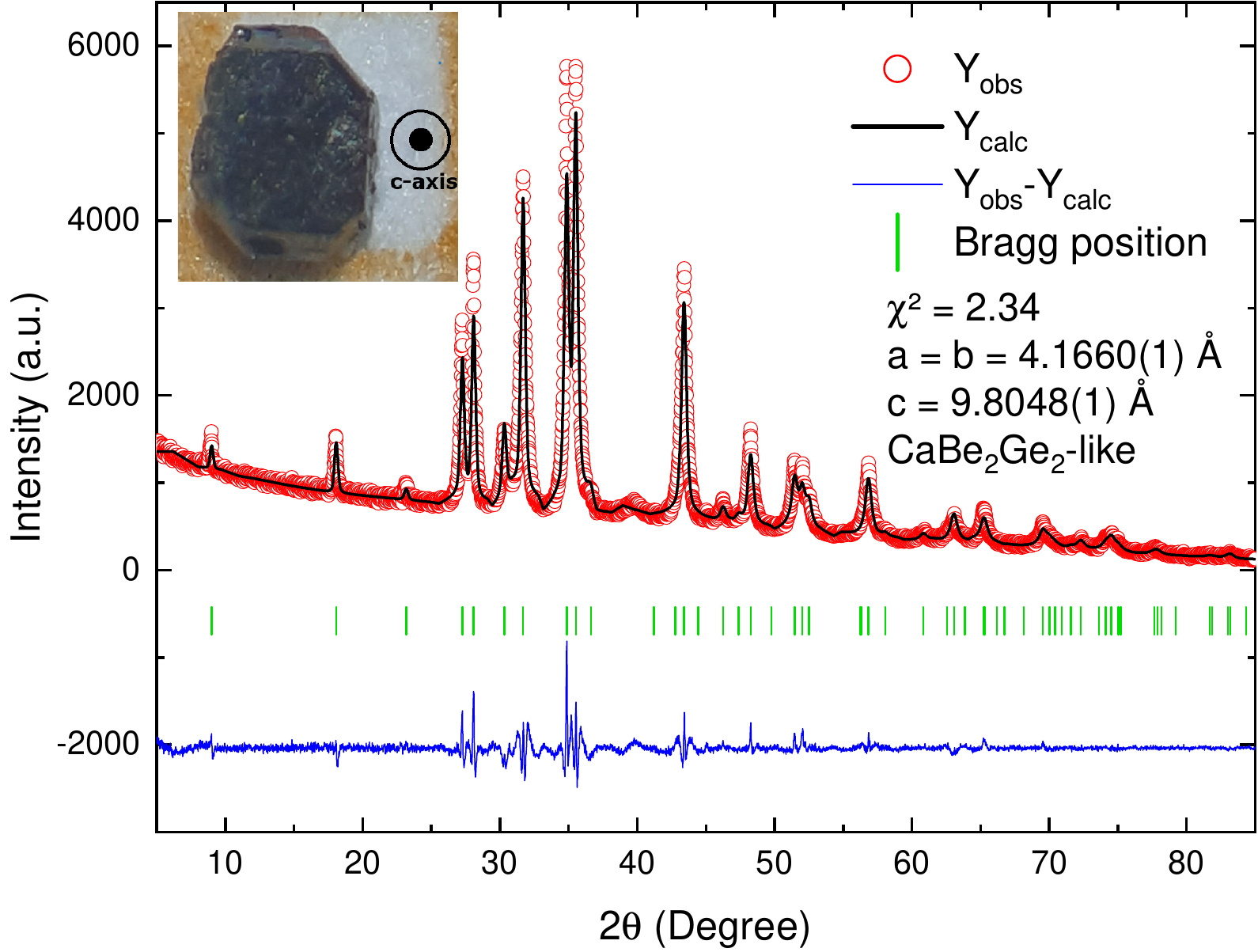}
\caption{\label{fig:DRX} Powder XRD pattern at room temperature. The solid black lines represent the Rietveld refinement fit, the solid blue line represents the difference between the observed and calculated profile, and the vertical green lines shows the Bragg positions.}
\end{figure}

The X-ray diffraction pattern for GdPt$_2$Si$_2$ is shown in  Fig.~\ref{fig:DRX} together with the Rietveld refinement, which confirms crystallization in the \textit{P}4\textit{/nmm} space group (CaBe$_2$Ge$_2$-like structure).
The refined lattice parameters are $a=b=4.1660(1)~$\AA$~$and $c=9.8048(1)~$\AA$~$ (Table \ref{tab:XRD_parameters}), which deviate from those reported for polycrystals by less than 2~\% \cite{GIGNOUX1991,HIEBL1985,MAYER1977}.
No additional diffraction peaks were observed, indicating a single phase free of any detectable Sn. 
This was also corroborated by the EDX analysis, by which the approximate stoichiometry of Gd$_{1.1(1)}$Pt$_{1.9(5)}$Si$_{1.8(8)}$ was obtained without any detectable trace of Sn.

\begin{table}[t]
\centering

\begin{tabular}{c|c|ccc}
\hline
Lattice parameters & Atom site & \multicolumn{3}{c}{Position}                                 \\ \hline
\textit{P}4\textit{/nmm} (\#129)     &           & \multicolumn{1}{c|}{x}    & \multicolumn{1}{c|}{y}    & z      \\ \hline
$a = 4.1660(1)~$\AA     & Gd (2\textit{c})    & \multicolumn{1}{c|}{0.25} & \multicolumn{1}{c|}{0.25} & 0.7488(6)\\ \hline
$c = 9.8048(1)~$\AA     & Pt1 (2\textit{a})  & \multicolumn{1}{c|}{0.75} & \multicolumn{1}{c|}{0.25} & 0 \\ \hline
$V = 170.1(6)~$\AA$^3$      & Pt2 (2\textit{c})  & \multicolumn{1}{c|}{0.25} & \multicolumn{1}{c|}{0.25} & 0.3746(1)\\ \hline
                   & Si1 (2\textit{b})  & \multicolumn{1}{c|}{0.75} & \multicolumn{1}{c|}{0.25} & 0.5 \\ \hline
$\chi ^2 = 2.34$& Si2 (2\textit{c})  & \multicolumn{1}{c|}{0.25} & \multicolumn{1}{c|}{0.25} & 0.1125(7)\\ \hline
\end{tabular}

\caption{\label{tab:XRD_parameters}
Lattice parameters, unit cell volumes and atomic coordinates of GdPt$_2$Si$_2$ at 300$~$K.}    

\end{table}
\subsection{Magnetic measurements}

\begin{figure}[]
\centering
\includegraphics[scale=0.32]{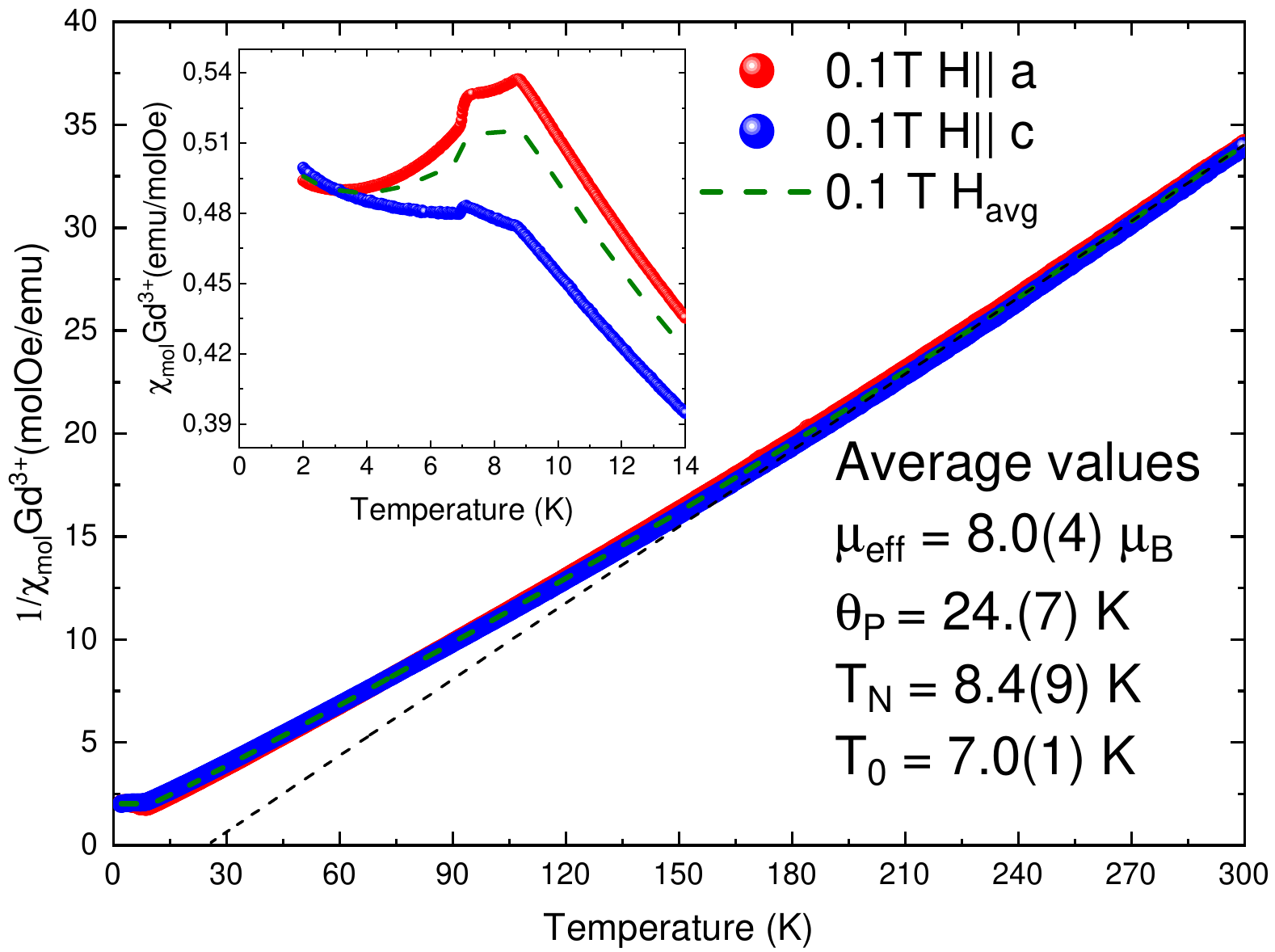}
\caption{\label{fig:sus} Inverse magnetic susceptibility for the main crystallographic directions (red and blue symbols). The black dotted line is a linear fit of the high-temperature region, from which the effective magnetic moment and the Weiss temperature were determined. The inset highlights the orientation dependence of the magnetic susceptibility ($\chi$) in the double magnetic transition region, plus the calculated polycrystalline average (green dashed line).}
\end{figure}

Fig.~\ref{fig:sus} displays the inverse magnetic susceptibility in an applied field of 0.1~T along both principal crystallographic directions and the polycrystalline average, calculated as $(2 \chi_a  + \chi_c)/3$.
At high temperatures there is no significant magnetic anisotropy, as expected for materials insensitive to CEF effects.
Above 190~K the response follows Curie-Weiss behavior, giving $\mu_{eff} = 8.0(4)~\mu_B$ (close to the expected Gd$^{3+}$ effective moment $\mu_{eff} = 7.94~\mu_B$) and positive Weiss paramagnetic temperature $\theta_p = 24.(7)$~K. 
This indicates a dominance of FM in-plane interactions over intra-plane AFM interactions in that temperature range, consistent with the reported Mössbauer spectra results \cite{NOWIK1980}. 
The same analysis below 120~K indicates that the AFM interactions become dominant and the Weiss temperature reaches $\theta_p = -6.6(5)$~K.

\begin{figure}[]
\label{fig:mag}
\includegraphics[scale=0.25]{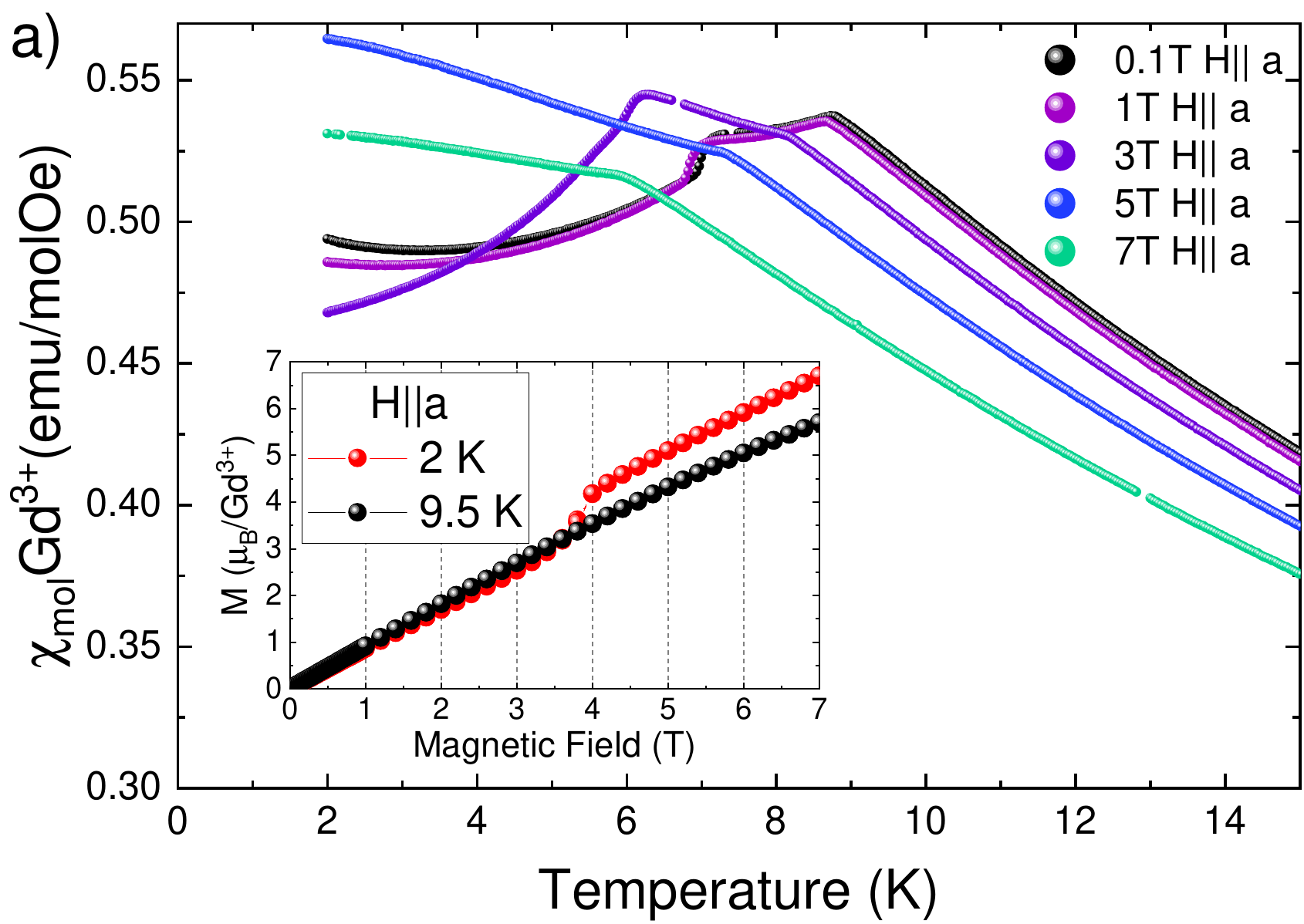}
\includegraphics[scale=0.25]{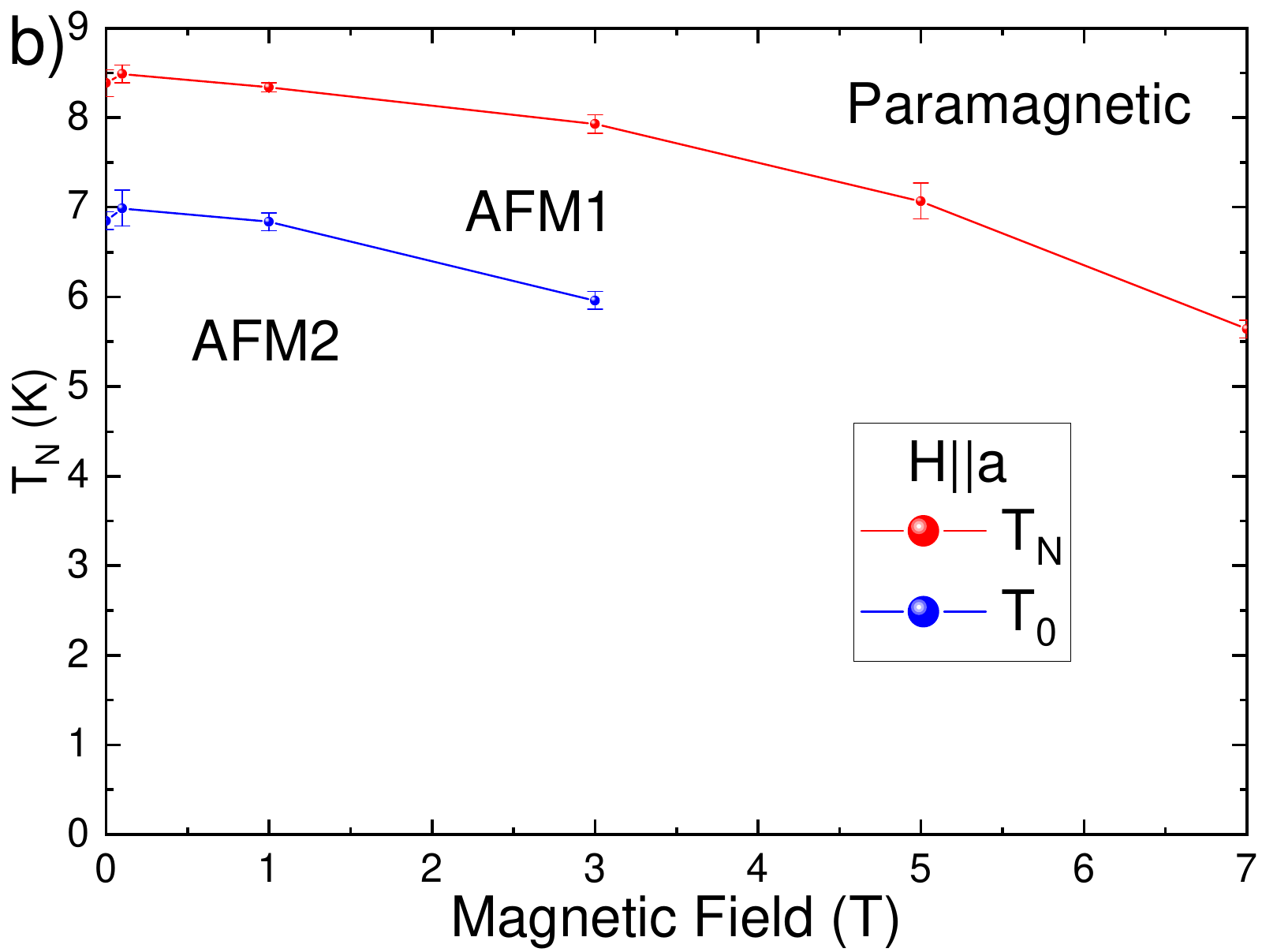}
\\
\includegraphics[scale=0.25]{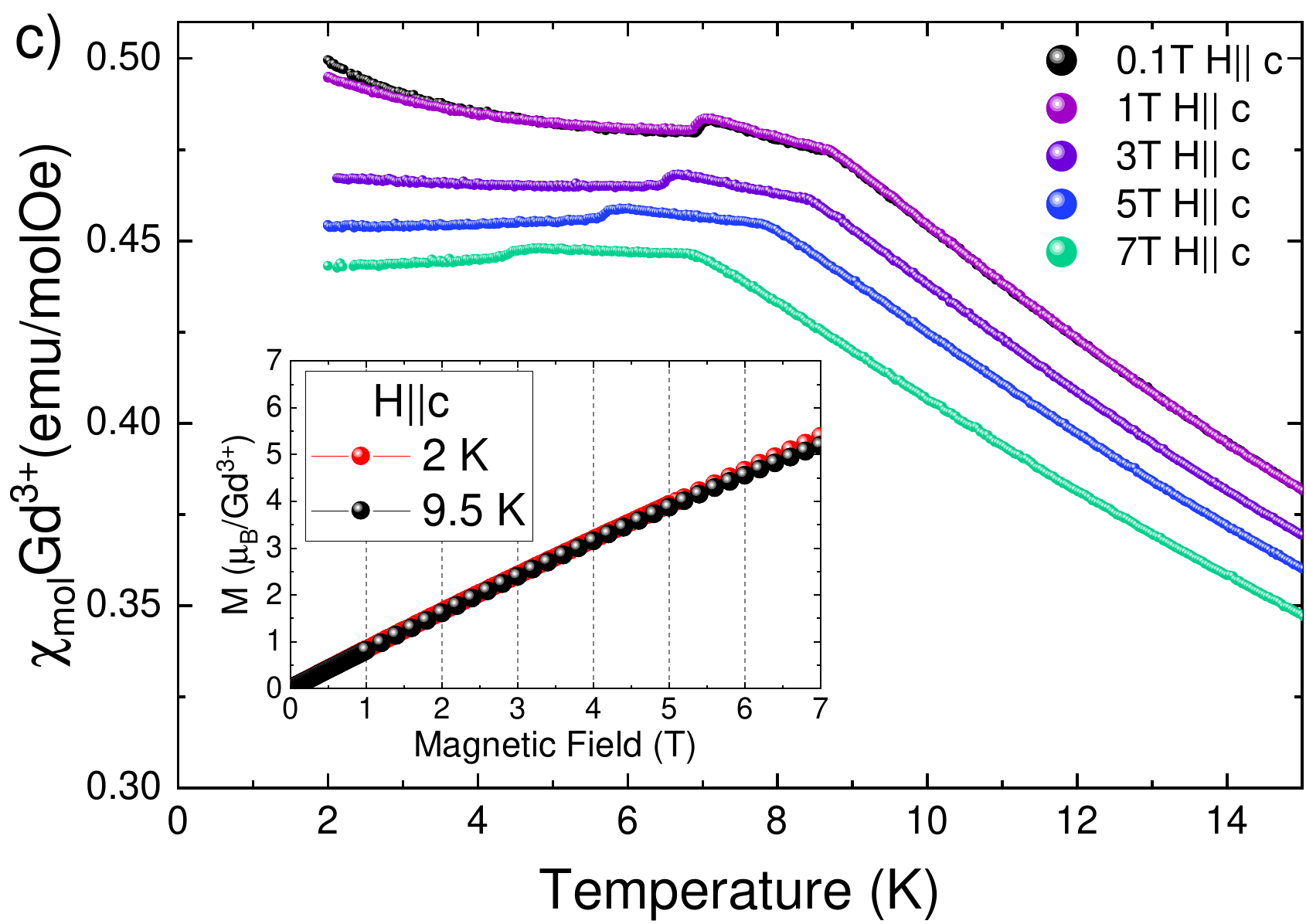}
\includegraphics[scale=0.25]{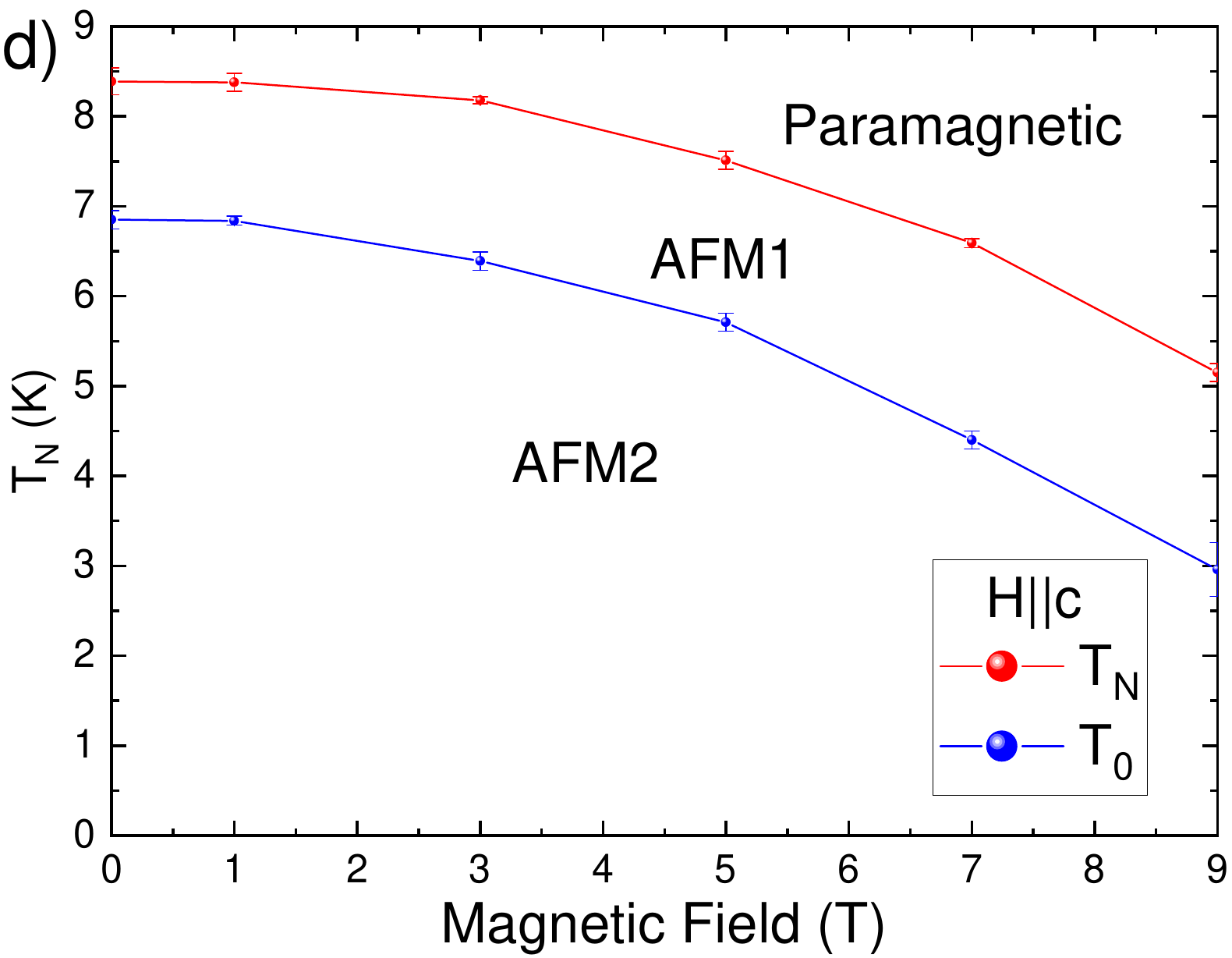}
\caption{\label{fig:mag} a) and c) displays the magnetic susceptibility as a function of temperature and in the insets the $M(H)~vs.~H$ for $H \parallel a$, and $H \parallel c$ respectively. b) and d) show the magnetic phase diagram for the two main crystallographic directions obtained by magnetic susceptibility and heat capacity measurements.}
\end{figure}

The inset of Fig. \ref{fig:sus} focuses on the magnetic susceptibility ($\chi$) in the region where the double magnetic transition lies. 
$T_N = 8.4(9)~$K and $T_0 = 7.0(1)~$K were determined by the maxima in $d(\chi T)/dT$ (not shown). 
The decrease in susceptibility below $T_0$ for $H \parallel a$ and the nearly constant response for $H \parallel c$ is typical behavior for AFM orderings without CEF anisotropy and magnetic moments aligned along the $a$-axis, in accordance with the proposed magnetic structure for GdPt$_2$Si$_2$\cite{GIGNOUX1991}.

The magnetic susceptibilities under various applied magnetic fields are displayed in Figs.~\ref{fig:mag}a) and ~\ref{fig:mag}c) for the region below 15~K and for the main crystallographic directions.
The insets in both figures show the respective $M~vs.~H$ for a temperature below $T_0$ and another above $T_N$.
When $H\parallel a$, a spin reorientation occurs, resulting in a metamagnetic transition that suppresses $T_0$. 
This behavior is not observed for $H\parallel c$, where the magnetic transitions shift to lower temperatures.

From our collective set of data, magnetic phase diagrams for $H \parallel a$ and $H \parallel c$ were constructed and are shown in Figs.~\ref{fig:mag}b) and ~\ref{fig:mag}d). 
They differ from the one proposed in Ref.~\cite{Duijn_2000MR}, since with polycrystalline samples $T_0$ is suppressed for 4~T, behavior observed only when $H \parallel a$ in our data.

\subsection{Electron Spin Resonance}

\begin{figure}
\centering
\includegraphics[scale=0.31]{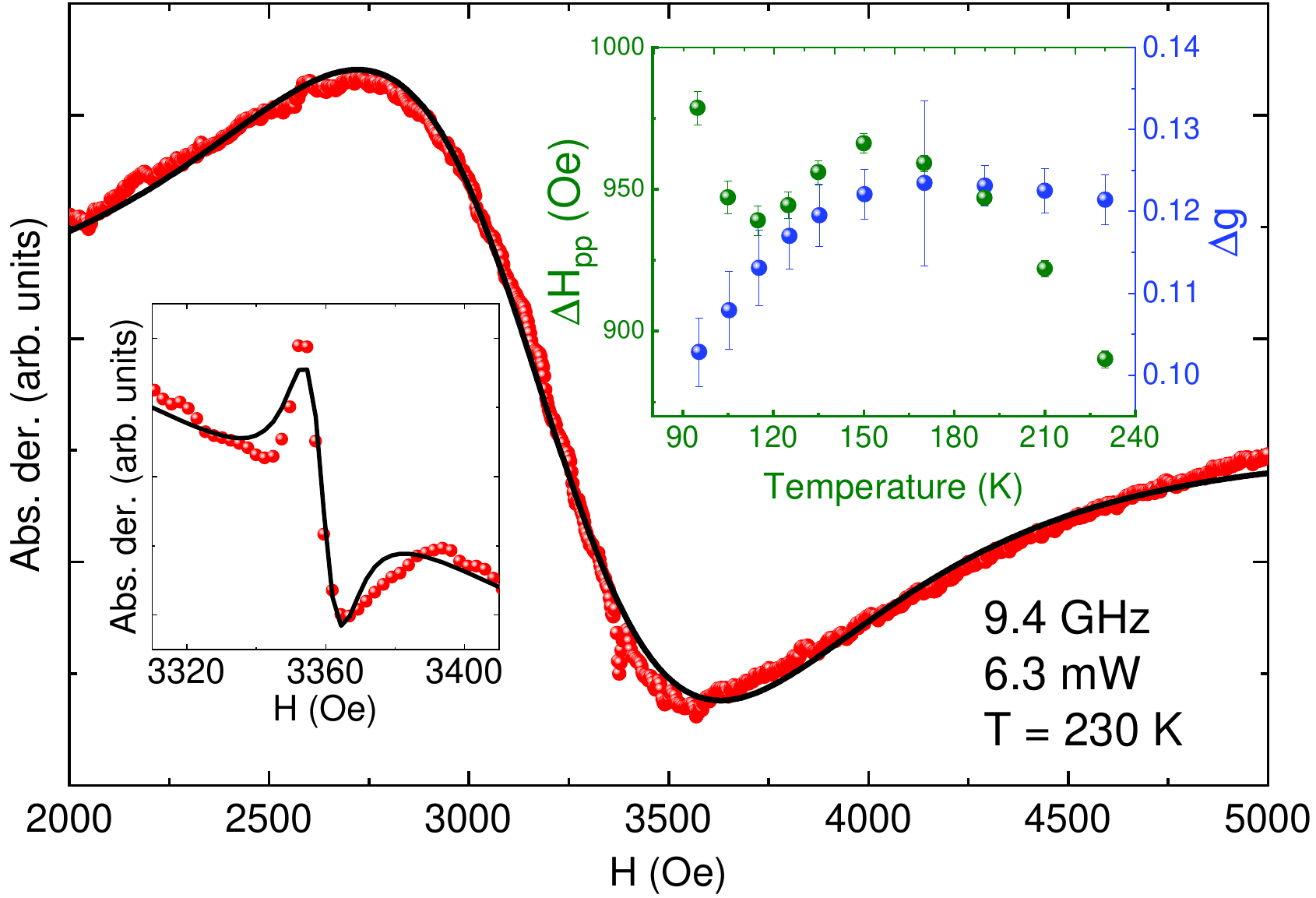}

\caption{\label{fig:esr}ESR spectra of Gd$^{3+}$ at $T=230$~K, $\nu = 9.4$~GHz and 6.3~mW (red symbols) together with the fitting using a Lorentzian equation (black line). The upper inset shows the temperature dependence of the linewidth and $g$-shift values obtained from the Lorentzian fits of the full spectra. The lower inset shows zooms in upon the region where a smaller resonance is observed.}
\end{figure}

The temperature-dependent ESR spectra for GdPt$_2$Si$_2$ (Fig.~\ref{fig:esr}) was obtained from single crystals crushed into fine powders of particle size smaller than 40~$\mu$m. 
In cases where the resonance originates from conduction electrons, it is expected that the linewidth follow the resistivity trends \cite{abragam1970electron}. 
For metallic systems the spectra is described by the Dyson theory. 
In the limiting case where the size of the particles ($d$) is larger than the skin depth ($\delta$) the spectra can be described by a Lorentzian lineshape \cite{esr_1_1955,esr_2_1955}, as is the present case since $d < 40~\mu$m and $\delta = \sqrt{2\rho/2\pi \nu \mu_0} \approx 16~\mu$m, where $\rho$ is the resistivity and $\mu_0$ the vacuum permeability. 
It is clear that the observed resonance is attributed to the Gd$^{3+}$ ion.
The Gd$^{3+}$ ESR spectra were thus fitted with a single Lorentzian resonance line.

\begin{equation}
  \frac{d \Delta H}{dH} \propto \frac{\Delta H_{pp}}{(\Delta H_{pp})^2 + (H - H_0)^2}\frac{}{}
\end{equation}

\noindent where $\Delta H_{pp}$ is the peak-to-peak resonance linewidth, $H$ is the applied magnetic field, and $H_0$ is the resonance field, used to calculate the $g$-factor by the relation $g = 71.444 \nu / 0.1 H_0$.

Shifts in the $g$-factor with respect to $g_0 = 1.993$ for Gd$^{3+}$ in insulators (used as reference \cite{abragam1970electron}) are associated with the presence of internal fields produced by the conduction electrons (\textit{ce}) surrounding the magnetic probe site.
Usually, positive/negative $g$-shifts ($\Delta g = g - g_0$) are attributed to FM/AFM interactions, respectively \cite{rosa2012electron}.
The sign of these shifts is directly related with the exchange interactions between the \textit{ce} spin and the $S$ spin lattice, which can provide information about the Fermi surface.
Note that spin-flip events are not expected to contribute to shifts of the $g$-factor ($\Delta g =0$).
The upper inset in Fig.~\ref{fig:esr} shows that the local polarization of the conduction electrons in GdPt$_2$Si$_2$ results in positive $g$-shifts.
As the system cools below 180~K $\Delta g$ decreases, indicating an increasing strength of the AFM interaction and/or an attenuation of FM interaction.

In metallic systems, it is typically expected that relaxation through \textit{ce} should lead to a linear broadening of the linewidth as the temperature increases, known as the Korringa rate \cite{abragam1970electron}. 
Our $\Delta H_{pp}$ analysis (same upper inset) does not reveal any indication of Korringa relaxation, at least within the measured range.
Instead, the line broadens for $T < 120$~K, which can be related to the onset of magnetic fluctuations at high temperatures. 
This hypothesis is further supported by the large deviation from the Curie-Weiss law in the $\chi(T)^{-1}$ presented in the magnetic data.

The lower inset in Fig.~\ref{fig:esr} highlights a smaller second resonance centered around $H = 3359$~Oe with a Lorentzian line shape. 
In contrast with the main resonance, this one has constant linewidth ($\Delta H = 10 \pm 1$~Oe), and $g$-factor ($1.998 \pm 0.009$).
Within error, the $g$ value is about the same as the Gd$^{3+}$ reference value.

\subsection{Specific Heat}

\begin{figure}
\includegraphics[scale=0.25]{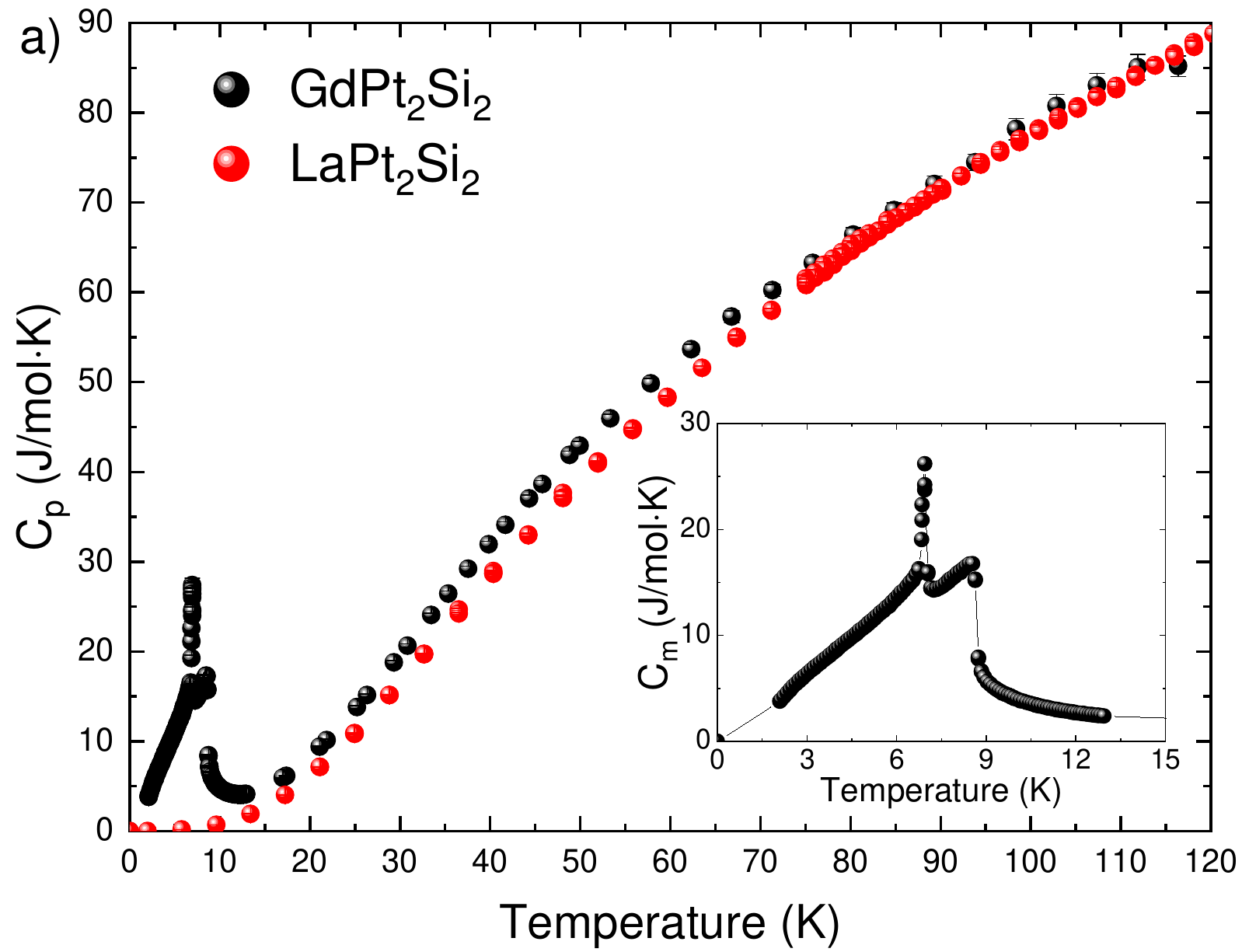}
\includegraphics[scale=0.25]{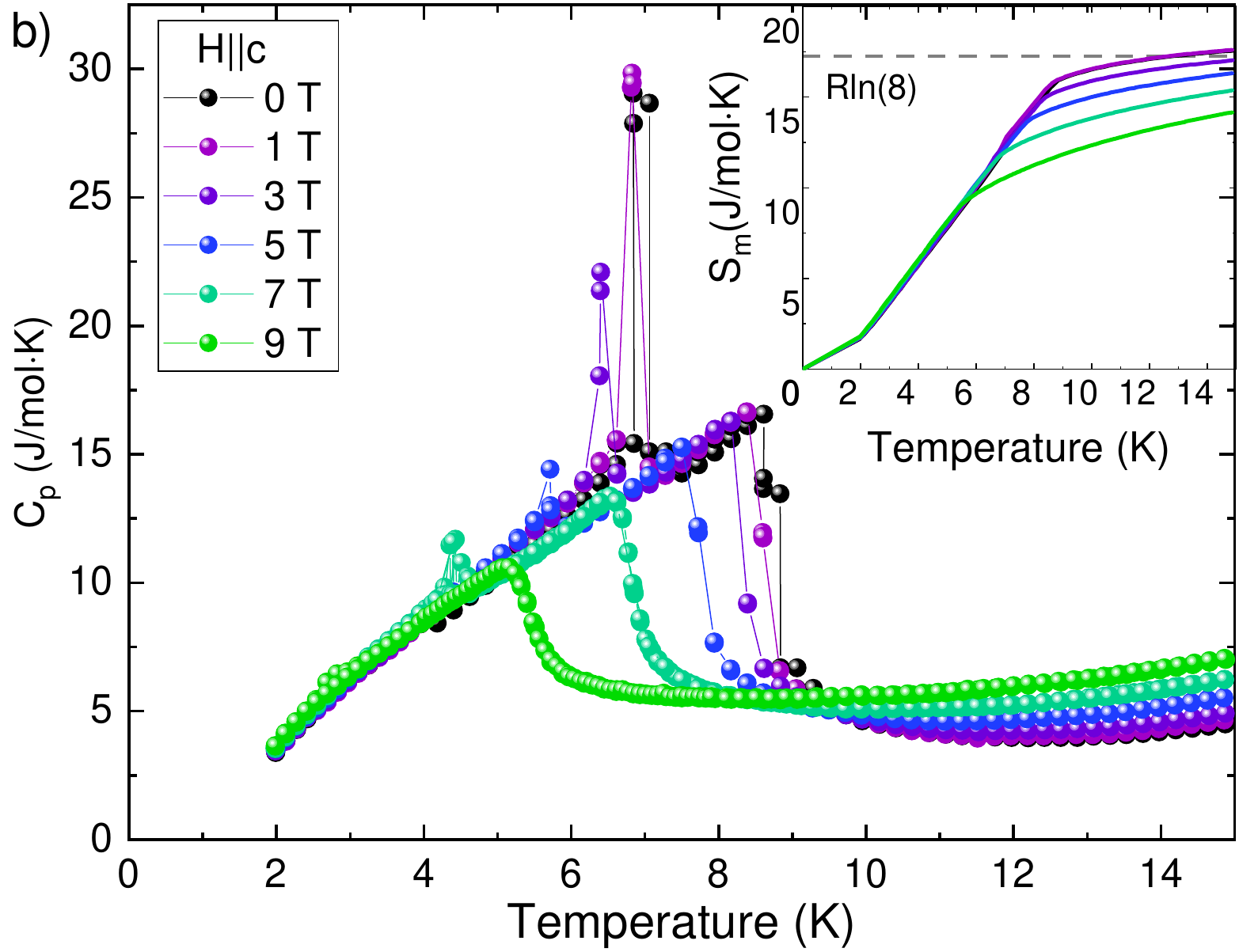}

\caption{\label{fig:cp} (a) Temperature-dependent specific heat measurements for single crystalline GdPt$_2$Si$_2$ and its non-magnetic reference LaPt$_2$Si$_2$. The inset displays the magnetic heat capacity ($C_m$) in the region of the double transition. (b) Total heat capacity up to 15 K for various applied external magnetic fields $H \parallel c$, revealing the suppression of the magnetic transition. The inset displays the magnetic entropy ($S_m$), where the expected value of Rln(8) is recovered right above $T_N$.}
\end{figure}

Fig.~\ref{fig:cp}a shows the specific heat data for GdPt$_2$Si$_2$ (black) and its non-magnetic counterpart, LaPt$_2$Si$_2$ (red). 
The two peaks displayed in the inset of the figure are related to magnetic order transitions, centered at $T_N = 8.5(4)$~K and $T_0 = 6.9(4)$~K, with $\Delta C_{T_N} \approx 13.9$~J/mol$\cdot$K and $\Delta C_{T0} \approx 21.8$~J/mol$\cdot$K respectively.
The mean-field approximation suggests that the first transition ($T_N$) is from a paramagnetic state to an amplitude modulated AFM one, below which at $T_0$ the system goes to an equal moment compensated magnetic structure \cite{Rotter_Cp_2001}. 
Upon increasing applied external magnetic field ($H\parallel c$), $T_0$ is rapidly suppressed, while $T_N$ shifts to lower temperatures (Fig.~\ref{fig:cp}b)).

The magnetic specific heat $C_m$ for each field was evaluated by subtracting the respective $C_p$ from LaPt$_2$Si$_2$. 
Numerical integration of $C_m/T$ was subsequently performed to obtain the magnetic entropies ($S_m$) displayed in the inset.
A linear extrapolation from 0 up to the lowest measured temperature was necessary to compensate the error introduced by the lack of data points below 2~K \cite{tari2003specific}. 
Since, in this material, the magnetism comes solely from the Gd$^{3+}$ $4f$ orbital, $S_m$ should reach its maximum value of $Rln(2J+1)=Rln(8) \approx 17.28$~J/mol$\cdot$K, when all $2J+1$ levels are populated, where $R=8.3145$~J/mol$\cdot$K is the gas constant.  Approximately 75\% of the magnetic entropy is reached right after $T_0$, and the maximum expected value is obtained for temperatures higher than $T_N$, indicating that an equal moment magnetic structure represents the ground state in this material due to the fact that no other transition is expected below 2 K.

\subsection{Electrical transport}

\begin{figure}
\centering
\includegraphics[scale=0.33]{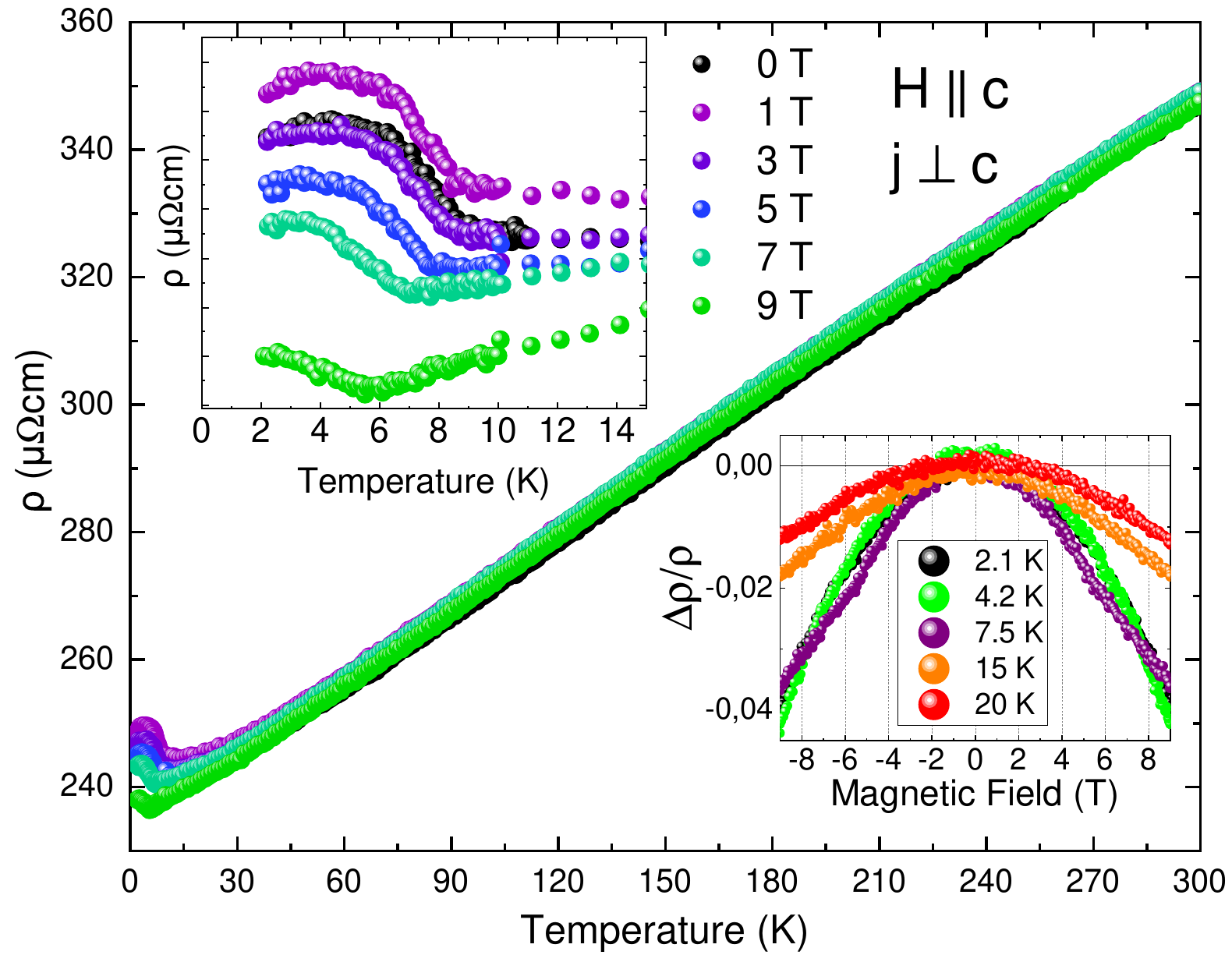}

\caption{\label{fig:rho}Electrical resistivity as a function of temperature for various applied magnetic fields $H \parallel c$, with $j \perp c$. The upper inset highlights the upturn related to the AFM transition, as well as its suppression with increasing fields. The lower inset displays the magnetoresistance at different temperatures.}
\end{figure}

Electrical transport measurements were carried out for $j \perp c$, showing metallic behavior with linear temperature-dependent resistivity at temperatures above 50~K, similar to other GdT$_2$Si$_2$ and RPt$_2$Si$_2$ compounds \cite{Malik_1998, Fushiya_2016}. 
Unlike the report on polycrystalline samples, there is an upturn of $\rho$ below $T_N = 8.6(7)~$K, but there is no indication of the second magnetic transition (Fig.~\ref{fig:rho}). 
The increase of resistivity in the magnetically ordered state can be interpreted as a modification of the Fermi surface, and since the periodicity of the lattice becomes larger when the system is in the AFM phase, fewer states contribute to the electrical conductivity. 
The magnetic moments align ferromagnetically in each $ab$-plane and couple antiferromagnetically between adjacent planes, similarly to the other layered tetragonal compounds such as GdRh$_2$Si$_2$ \cite{KLIEMT201537, Güttler2016, Duijn_2000MR}.

Under applied magnetic field along the $c$-axis, the spin fluctuations have small influence at high temperatures and the linear behavior extends down to $\sim 10$~K. 
Taking this minimum into account, the estimated residual resistivity ratio $RRR=\rho_{300K}/\rho_{10K} = 1.47$ is comparable to that of the nonmagnetic comopunds of this series, suggesting a tendency of crystallographic disorder \cite{Nagano2013, meaden2013electrical}. 
For all $\rho \times T$ measurements, the upturn shifts to lower temperatures with increasing fields, together with a reduction of resistivity, implying a negative magnetoresistance ($MR$).
The $H = 1$~T curve is an exception, where the resistivity is enhanced (upper inset in Fig.~\ref{fig:rho}). 
In the MR isotherms (lower inset in Fig.~\ref{fig:rho}) positive values are clearly observed only at $T = 4.2~$K, which may be related to spin fluctuations. 
The isotherms show quadratic MR behavior with remarkable deviation above $H = 4~$T for $T = 7.5~$K, which matches the region of the metamagnetic transition found in $M~vs.~ H$ curves for $H \parallel a$, indicating that it is related to spin reorientation.

\section{Summary}

Single crystals of GdPt$_2$Si$_2$ were successfully grown by Sn flux, resulting in mm-sized crystals. 
The specific heat data revealed a pair of magnetic transitions at $T_N = 8.5(4)~$K and $T_0 = 6.9(4)~$K, the former being lower than the reported value for polycrystalline samples ($\approx 10~$K) and the latter at the same temperature. 
In resistivity measurements, the material showed metallic behavior and only $T_N$ could be observed. 
The low RRR reveals the tendency of crystallographic disorder in this series. 
The effective moment $\mu_{eff} = 8.0(4)~\mu_B$, obtained from magnetic measurements, agrees with the theoretically predicted value for Gd$^{3+}$, and the positive Weiss temperature ($\theta_p = 24.(7)~$K) indicates a pronounced competition between AFM and FM interactions, corroborated by the ESR measurements.

The magnetotransport anomaly around 1~T is being further investigated by Hall measurements and orientation-dependent fields. 
Despite the indication that GdPt$_2$Si$_2$ has an amplitude-modulated AFM order, there is a possibility of a complex noncollinear AFM structure \cite{lkatka1988magnetism} that emerges from antisymmetric exchange interactions, in that sense, further work aims at in-depth understanding its magnetic structure, and how the break of the inversion symmetry influences the magnetic ordering compared to the other GdT$_2$X$_2$.
Such understanding will be useful as a platform to help comprehend the responses of the other rare earth members in the RPt$_2$Si$_2$ series (R = rare earth) \cite{Fushiya_2016, NdPt2Si2_2020, SmPt2Si2_2021, EuPt2Si2_2018}.

\section*{Acknowledgments}
We acknowledge the financial support of Brazilian funding agencies CAPES, CNPq (Contracts No. 140921/2022-2, No. 88887.837417/2023-00), FAPESP (No. 2017/20989-8, No. 2017/10581-1), and the experimental support from Multiuser Central Facilities (UFABC) and LCCEM (UFABC). We are thankful to M. H. Carvalho da Costa for assistance in the experiments conducted at GPOMS/UNICAMP.


\bibliographystyle{elsarticle-num} 
\bibliography{bibl}





\end{document}